\documentclass[aps,prc,twocolumn,showpacs,eqsecnum,superscriptaddress,floatfix,nofootinbib]{revtex4-1}
\usepackage{amsmath,graphicx}
\usepackage{color}

\makeatletter
\newcommand\erfc{\mathop{\operator@font erfc}\nolimits}
\def\slashchar#1{\setbox0=\hbox{$#1$}
   \dimen0=\wd0 \setbox1=\hbox{/} \dimen1=\wd1
   \ifdim\dimen0>\dimen1 \rlap{\hbox to \dimen0{\hfil/\hfil}} #1
   \else  \rlap{\hbox to \dimen1{\hfil$#1$\hfil}} / \fi}

\makeatother

\begin{document}
\title{Femtoscopy analysis of d-Au interactions at $\sqrt{s}=200$GeV}
\author{Piotr Bo\.zek}
\email{Piotr.Bozek@ifj.edu.pl}
\affiliation{AGH University of Science and Technology, Faculty of Physics and Applied Computer Science, al. Mickiewicza 30, PL-30059 Krakow, Poland}
\affiliation{The H. Niewodnicza\'nski Institute of Nuclear Physics, Polish Academy of Sciences, PL-31342 Krak\'ow, Poland}

\date{\today}

\begin{abstract}
The femtoscopy correlation radii  for d-Au
 collisions at $200$GeV are calculated in the  hydrodynamic model and 
compared to PHENIX collaboration data. For asymmetric systems, such as d-Au or p-Pb collisions, 
 the   correlation radius   $R_{out-long}$ is estimated and   predicted to be non-zero. It appears due to different lifetimes
of the parts of the fireball at positive and negative rapidities. 
Azimuthally sensitive 
Hanbury Brown-Twiss analysis with respect to the second order event 
plane shows a significant angular dependence of the radii. 
It  reflects the strong azimuthal asymmetry
of the d-Au source geometry and flow for central interactions.
\end{abstract}



\maketitle

\section{Introduction}

The dynamics of ultrarelativistic collisions between a large nucleus and a  smaller projectile 
is the subject of intensive experimental and theoretical studies \cite {Sickles:2014dca,*Bozek:2014era,*Venugopalan:2014tla}.
One of the key characteristic of the system is 
the size of the interaction region \cite{Bozek:2013df,Bzdak:2013zma,Abelev:2014pja,*cmswikihbt}.
The PHENIX collaboration published results on the Hanbury Brown-Twiss (HBT) correlations for d-Au collisions 
at $200$GeV  \cite{Adare:2014vri}.
The results are for  three HBT radii  side  $R_{side}$, out $R_{out}$, and long $R_{long}$ ($R_{s,o,l}$) 
as function of the pion pair momentum and for 
different collision centralities.
Previously the PHENIX experiment presented results indicating  collective flow effects in the dynamics of d-Au 
collisions \cite{Adare:2013piz}. The density of matter created in central d-Au interactions is similar as in Au-Au collisions, which implies a collective expansion stage. Combined with 
 the large eccentricity of the fireball  in d-Au collisions it leads 
to a significant elliptic flow  of emitted particles 
\cite{Bozek:2011if}. In the hydrodynamic model  
hadrons
 are emitted from the  freeze-out hypersurface and the femtoscopy radii measure the size of
 the effective  emission region (the homogeneity region). The collective expansion 
of matter leads 
to strong space-momentum correlations for the emitted particles. The size of
 the homogeneity region measured with pion interferometry diminishes with increasing average 
pion pair momentum \cite{Wiedemann:1999qn,*Heinz:1999rw,*Lisa:2005dd}.

In central collisions of asymmetric systems, such as  d-Au and p-Pb  collisions, 
the charged particle density in rapidity 
is far from boost invariance and is not
symmetric in the central rapidity region \cite{Nouicer:2004ke,ATLAS-CONF-2013-096}.
The fireball formed in the collision is asymmetric in space-time rapidity, 
on the side of the larger nucleus it lives longer. In the following, it is shown that 
 an additional radius parameter appears 
in the  Gaussian form of the interferometry
correlations function,  representing a mixed out-long correlations term. The value of the
$R^2_{out-long}$ ($R^2_{ol}$)  parameter is predicted for d-Au and p-Pb collisions.

The azimuthal asymmetry of the emitting source can be  observed using the method of azimuthally sensitive  HBT 
correlations  \cite{Voloshin:1995mc,*Wiedemann:1997cr,*Lisa:2000ip,*Heinz:2002sq,Heinz:2002au}. 
In particular, the spatial eccentricity should be visible as an increase of the side
radius extracted for pion pairs emitted in plane. The azimuthal dependence of the interferometry 
radii with respect to the second order event plane 
involves even harmonics in its Fourier decomposition. In Sect. \ref{sec:ahbt}, it is shown that for  a system with strong quadrupole 
deformation, as in  d-Au collisions, the second harmonic component can be identified for all the three
HBT radii.

\section{HBT radii}

\label{sec:hbtr}

The dynamics of  d-Au and Au-Au collisions is described using the event-by-event $3+1$-dimensional viscous hydrodynamic model 
\cite{Bozek:2011ua}.
The initial density of the fireball formed in the collision is modeled
 using the Glauber Monte Carlo model
\cite{Rybczynski:2013yba}. For each initial density profile, the expansion is 
followed using viscous hydrodynamics with shear and and bulk viscosity. 
The  viscosity to entropy ratio is fixed at $0.08$ for the shear and $0.04$ for 
bulk viscosity \cite{Bozek:2009dw}.
At the freeze-out temperature of $150$MeV the collective expansion 
stops and particles are emitted, with subsequent resonance decays \cite{Rybczynski:2013yba}. 
Due to limited statistics that can be obtained in the simulations, for each hydrodynamic 
freeze-out hypersurface 5000 real events are generated and combined together to increase statistics. 
It means that the number of pion pairs per event is effectively increased $5000$ times.

\begin{figure*}
\includegraphics[width=.66\textwidth]{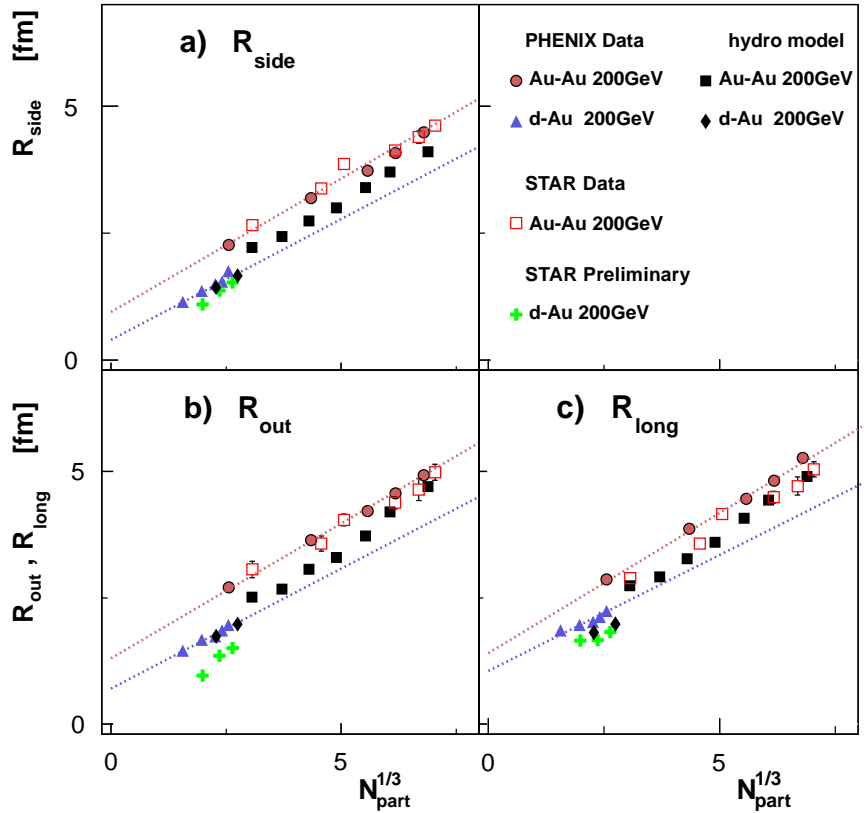}
\caption{The HBT radii $R_{side}$ (panel a)), $R_{out}$ (panel b)), and $R_{long}$ (panel c)) for d-Au and Au-Au collisions
at $200$GeV for different centralities (plotted as function of $N_{part}^{1/3}$). The experimental data of the PHENIX 
collaboration are represented using circles (Au-Au) \cite{Adler:2004rq} and up-triangles (d-Au) \cite{Adare:2014vri}, the Au-Au data of the STAR collaboration 
using open squares \cite{Adams:2004yc}, and the preliminary d-Au data of the STAR collaboration using crosses \cite{Chajecki:2009afa}. The results of the hydrodynamic model are shown using diamonds (d-Au) and solid squares (Au-Au \cite{Bozek:2014hwa}).}
\label{fig:hbt}
\end{figure*}

The correlation function $C(q,k)$ for a pion pair of average momentum $k$ and relative momentum $q$ is approximated by a histogram $C(q,k)$ in a  bin $q_a$, $k_b$  \cite{Bozek:2014hwa}
\begin{eqnarray}
C(q_a,k_b) =  &&\nonumber \\
\frac{\frac{1}{N_{pairs,num}}\sum_{j=1}^{N_h} \sum_{m, l=1  }^{N_e}\sum_{s=1}^{M_l}\sum_{f=1 }^{M_m} \delta_{q_a}\delta_{k_b} \Psi(q,x_1-x_2)}
{\frac{1}{N_{pairs,den}}\sum_{i\neq j=1}^{N_h} \sum_{l,m=1  }^{N_e}\sum_{s=1}^{M_l}\sum_{f=1}^{M_m} \delta_{q_a}\delta_{k_b}} &&  \nonumber \\
&& . \label{eq:bincq}
\end{eqnarray}
The numerator is constructed summing over events $l$ and $m$ generated from the same freeze-out hypersurface $j$
( $M_l$ and $M_m$ are the  multiplicities of the respective events),
in the denominator the two events are generated from two different hydrodynamic events  $i$ and $j$.
If the relative momentum
 $q=p_s-p_f$ and the  average pair momentum $k=(p_s+p_f)/2$ fall into the respective bins, $\delta_{q_a}$ and $\delta_{k_b}$ are $1$ and $0$ otherwise.  $\Psi(q,x_1-x_2)=(e^{i q (x_1-x_2)}+ e^{-iq(x_1-x_2)})/\sqrt{2}$ is the symmetrized wave-function of the pion pair. Final state interactions between pions, are not taken into account in the above formula and no corrections
for such interactions are performed when fitting the  Gaussian formula to the correlation function.
The correlation function is constructed in the longitudinal comoving  system for pion pairs with rapidity $|y|<1$.
The details of the procedure for the hydrodynamic evolution and for the construction of the 
HBT correlation function in the event-by-event hydrodynamic model can be found in \cite{Bozek:2013df,Bozek:2014hwa}.

In a given average transverse pair momentum $k_\perp$  bin, the dependence of the
 correlation function $C(q,k_\perp)$ on the relative momentum $q$ is decomposed into  three components, 
$q_{l}$ along the beam axis, $q_{o}$ along the pair transverse momentum $k_\perp$ in the local comoving system, and $q_{s}$ orthogonal to the first two directions.
The Bertsch-Pratt formula is fitted in 3-dimensions 
\begin{equation}
C(q,k_\perp)=1+ \lambda  e^{-R_o^2 q_{o}^2-R_s^2q_{s}^2-R_l^2q_{l}^2} 
\ , \label{eq:gauss1} 
\end{equation}
where the extracted parameters $R_{o,s,l}$ are the three HBT radii 
\cite{Bertsch:1989vn,*Pratt:1986ev}.

In Fig. \ref{fig:hbt} is shown the dependence of the HBT radii on the average number of participant 
nucleons for the given centrality class, for $350$MeV$<k_\perp<450$MeV. The calculation reproduces rather well the experimental data 
of the PHENIX collaboration
 for d-Au collisions. The preliminary STAR collaboration results for $R_{o}$ in d-Au are below the PHENIX results.
The systematic uncertainty of the fitted results for the hydrodynamic calculation is estimated by varying the range of the fit from $|q|<0.1$GeV to $0.2$GeV, the variation  of the extracted radii is smaller than 10\%.

The calculated radii for  Au-Au collisions are systematically below the experimental values. The deviation is not reduced 
by lowering the freeze-out temperature to $140$MeV. % or increasing the width of Gaussian representing in the transverse plane re
The result shows that the hydrodynamic model can describe only partly the dependence of the 
HBT radii on the system size and the centrality of the collisions. It may indicate that the flow profile generated in 
the hydrodynamic simulation with Glauber model  initial conditions is only approximate or that 
contributions of the pre-equilibrium 
flow or rescattering stage are noticeable \cite{Broniowski:2008vp,*Pratt:2008qv,*Karpenko:2012yf}. 
Other effects could be also be important for the interferometry in small systems
 \cite{Sinyukov:2013zna,*Tomasik:2007gs,*Bialas:2014gca}.

\section{Azimuthally sensitive HBT correlation function in asymmetric collisions}

In this section I recall the basic formulae of the  azimuthally sensitive HBT analysis \cite{Voloshin:1995mc} and apply 
them to the correlation function with respect to the second order event-plane, for a system without 
forward-backward symmetry in rapidity.
The azimuthal  dependence of the HBT correlation function comes from the 
explicit dependence on the angle $\Phi$ 
between the emitted pion pair with respect to the event-plane and from the 
implicit dependence of the emission function $S(x,k)$ on the angle  
\cite{Voloshin:1995mc,Heinz:2002au}.
For a general Gaussian parametrization of the correlation function
\begin{equation}
C(q,k)=1+\lambda e^{-\sum_{i,j}R^2_{i,j}q_{i}q_{j}} \ \  , \ \  i=\mbox{o},\ \mbox{s},\ \mbox{l} \ \ ,
\end{equation}
the radii parameters $R^2_{i,j}$ are related to moments of the emission  function $S(x,k)$,
$S_{\mu\nu}=\langle x_\mu x_\nu \rangle-\langle x_\mu \rangle \langle x_\nu\rangle$, $z$ is along the beam direction, ($x$-$y$) is the transverse plane, the second order flow direction is $x$.
The symmetries of the emission function $S(x,k)$ can be used to infer the form the azimuthal dependence of the HBT radii \cite{Heinz:2002au}.
In central p-Pb and d-Au collisions the  fireball exhibits elliptic  and triangular deformation. 
For p-Pb collisions it occurs solely due to fluctuations, 
while for d-Au collisions, even a zero impact parameter, the intrinsic deformation of the deuteron
 dominates the fireball  eccentricity. In any case, the 
azimuthal angle is defined with respect to second order event plane in each event. 
In the average over many events, while keeping always the orientation with respect to the second order plane,
   one is left with  two symmetries of the emission source
\begin{eqnarray}
S(x,y,z,k_\perp,\Phi)&=&S(x,-y,z,k_\perp,-\Phi) \nonumber \\
S(x,y,z,k_\perp,\Phi)&=&S(-x,y,z,k_\perp,\pi-\Phi) \ .
\label{eq:symm}
\end{eqnarray}  
The Fourier expansion of the emission source moments respecting the symmetry is
\begin{eqnarray}
\frac{1}{2}\left((\langle x^2\rangle - \langle x \rangle^2 ) \right. & +& \left. (\langle y^2 \rangle
 - \langle y \rangle^2)\right)  = \nonumber \\ & & 
 A_0  +2\sum_{n=2,4,\dots} A_n \cos(n\Phi) \nonumber \\
\frac{1}{2}\left((\langle x^2\rangle - \langle x \rangle^2 ) \right. & - &  \left. (\langle y^2 \rangle
 - \langle y \rangle^2)\right)  =  \nonumber \\ & &  B_0  +2\sum_{n=2,4,\dots} B_n \cos(n\Phi) 
\nonumber \\
\langle x y \rangle - \langle x \rangle \langle y \rangle & =&   2\sum_{n=2,4,\dots} C_n \sin(n\Phi) 
\nonumber \\
\langle t^2 \rangle - \langle t \rangle^2 & =& D_0 + 2\sum_{n=2,4,\dots} D_n \cos(n\Phi) \nonumber \\ 
\langle t x \rangle - \langle t \rangle \langle x \rangle & =&   2\sum_{n=1,3,\dots} E_n \cos(n\Phi) 
\nonumber \\
\langle t y \rangle - \langle t \rangle \langle y \rangle & =&   2\sum_{n=1,3,\dots} F_n \sin(n\Phi) 
\nonumber \\
\langle t z \rangle - \langle t \rangle \langle z \rangle & =&  G_0 + 2\sum_{n=2,4,\dots} G_n \cos(n\Phi) 
\nonumber \\
\langle x z \rangle - \langle x \rangle \langle z \rangle & =&   2\sum_{n=1,3,\dots} H_n \cos(n\Phi) 
\nonumber \\
\langle y z \rangle - \langle y \rangle \langle z \rangle & =&   2\sum_{n=1,3,\dots} I_n \sin(n\Phi) 
\nonumber \\
\langle z^2 \rangle - \langle z \rangle^2 & =& J_0 + 2\sum_{n=2,4,\dots} J_n \cos(n\Phi) 
\end{eqnarray}
The HBT parameters $R^2_{i,j}(\Phi)$ of the Gaussian correlation 
function are obtained after rotation to the direction of the pair 
\cite{Wiedemann:1999qn,Heinz:2002au}. The first terms of the Fourier expansion of the azimuthal 
dependence of the HBT radii are
\begin{eqnarray}
R^2_{s} & = & A_0 - B_2 -C_2 + (2A_2-B_0-B_4-C4) \cos(2\Phi) \nonumber \\
R^2_{o} & = & A_0 + B_2 + C_2 -2 E_1 \beta_\perp -2 F_1\beta_\perp + D_0 \beta_\perp^2 \nonumber \\
&+& (2A_2+B_0+B_4+C4 -2\beta_\perp(E_1+E_3-F_1+F_3) \nonumber \\
 & & +2 D_2 \beta_\perp^2) \cos(2\Phi) \nonumber \\
R^2_{os} & = & \left( -B_0 + B_4 +C_4 + \beta_\perp(E_1-E_3-F_1-F_3) \right) \nonumber \\
 & & \sin(2\Phi) \nonumber \\
R^2_{l} & = & J_0  + 2 J_2 \cos(2\Phi) \nonumber \\
R^2_{ol} & = & H_1 + I_1   -  G_0 \beta_\perp  + (I_1+I_3-H_1+H_3) \cos(2\Phi) \nonumber \\
R^2_{sl}& =& (I_1+I_3-H_1+H_3) \sin(2\Phi) \ ,
\label{eq:aradii}
\end{eqnarray}
where $\beta_\perp=k_\perp/k_0$. The symmetry does not constraint the angle independent term of $R^2_{ol}$ to vanish.

\section{$R_{out-long}$ in asymmetric collisions}

\label{sec:rol}

\begin{figure}
\includegraphics[width=.43\textwidth]{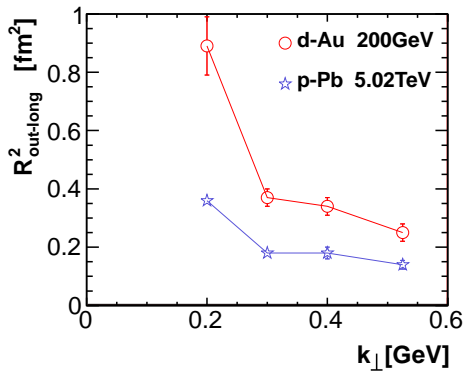}
\caption{The HBT radius $R_{ol}^2$ for d-Au collisions at $200$GeV (circles) and for p-Pb collisions
at $5.02$TeV (stars) as function of the pion pair momentum. }
\label{fig:rol}
\end{figure}

In asymmetric collision one  finds four angle independent radii parameters (Eq. \ref{eq:aradii}) 
and the HBT correlation function has the  form
\begin{equation}
C(q,k_\perp)=1+ \lambda  e^{-R_o^2 q_{o}^2-R_s^2q_{s}^2-R_l^2q_{l}^2-2R_{ol}^2 q_{o}q_{l}} 
\ . \label{eq:gauss2} 
\end{equation}
The radii are fitted to the angle  averaged correlation function $C(q,k_\perp)$ obtained from the hydrodynamic simulations, the same as used in the standard HBT analysis in Sect. \ref{sec:hbtr}.
The three radii $R_{o}$, $R_{s}$, and $R_{l}$ are the same as obtained with the simpler formula (\ref{eq:gauss1}).
However, 
one finds an asymmetry of the correlation function in the out-long direction, both in d-Au collisions at RHIC and p-Pb collisions at the LHC, which results in a nonzero value of $R^2_{ol}$ (Fig. \ref{fig:rol}).

\begin{figure}
\includegraphics[width=.38\textwidth]{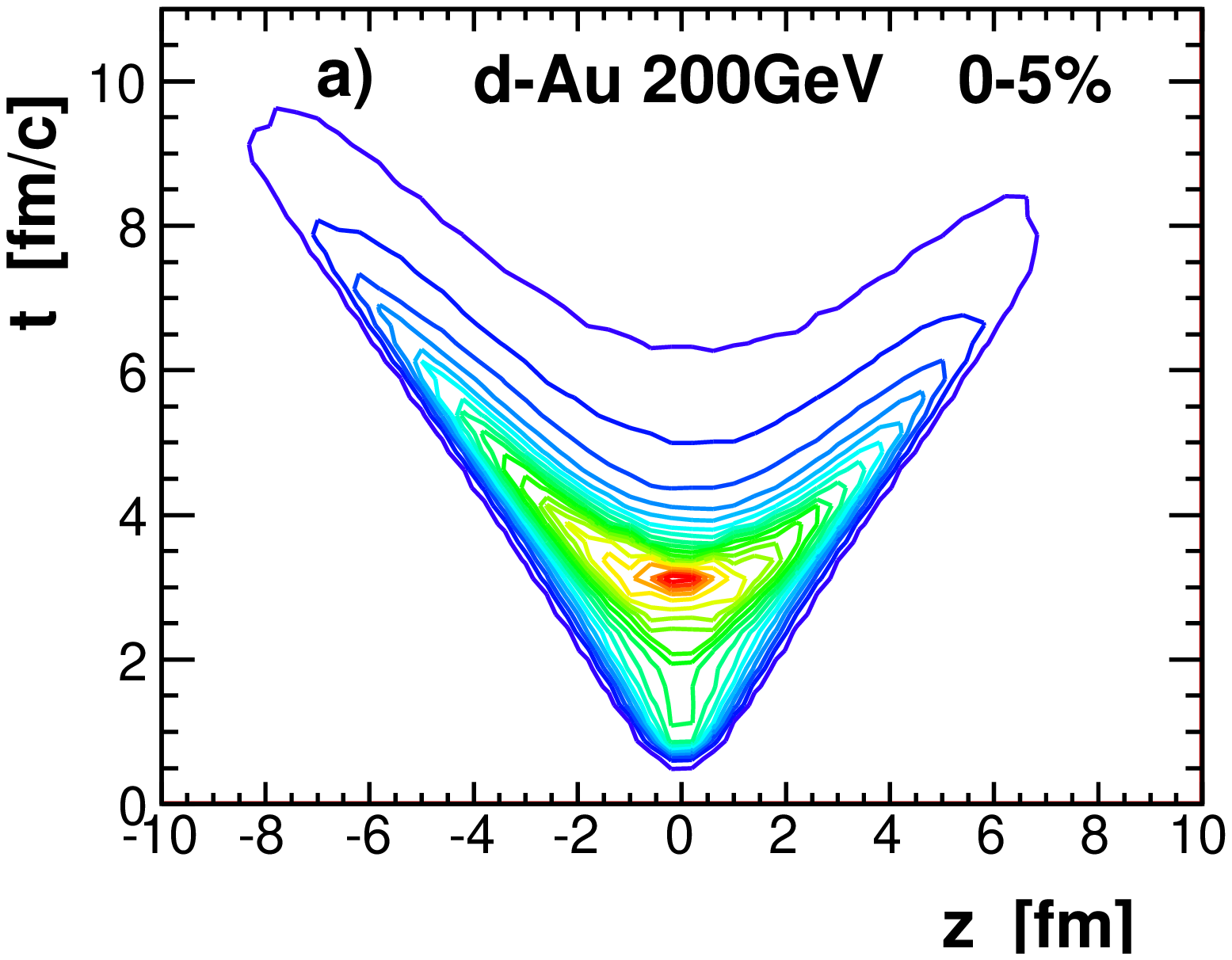}\\
\includegraphics[width=.38\textwidth]{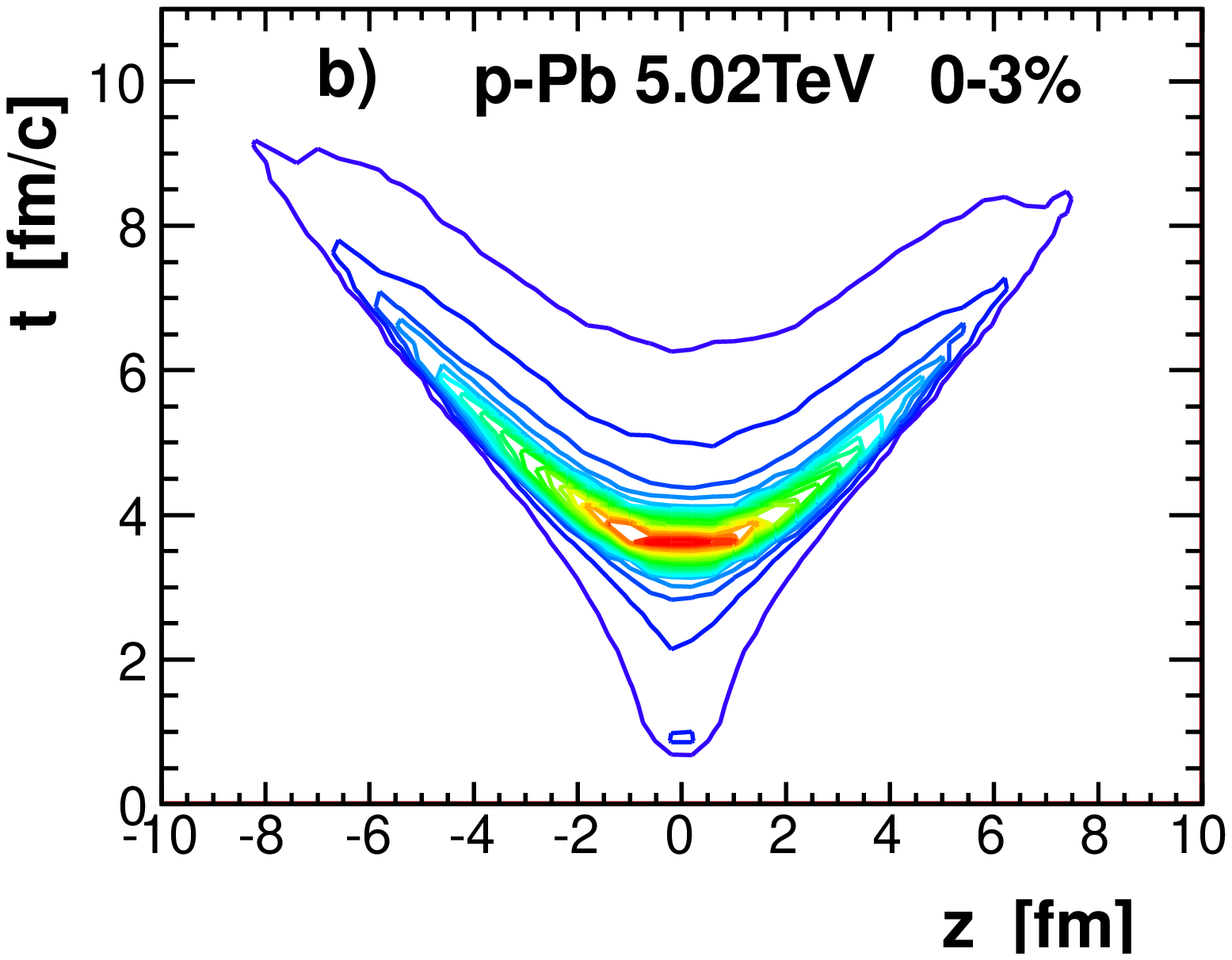}\\
\caption{The $(z,t)$ emission point  density of  charged pions emitted in  d-Au 
collisions at $200$GeV (panel a)
and p-Pb collisions at $5.02$TeV (panel b), in the longitudinal comoving frame of the pion. }
\label{fig:zt}
\end{figure}

\begin{figure}
\includegraphics[width=.38\textwidth]{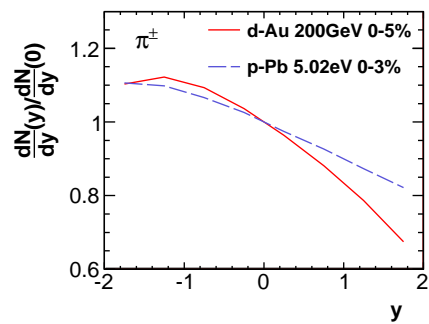}
\caption{The scaled rapidity distribution for charged pions in d-Au collisions at $200$GeV (solid line)
and p-Pb collisions at $5.02$TeV (dashed line).  }
\label{fig:dndy}
\end{figure}

The nonzero value of $R^2_{ol}$ for the correlation function in a symmetric window around midrapidity
indicates that the fireball freezes out asymmetrically at positive 
and negative rapidities. On the side where the larger projectile goes, the fireball lives longer 
(we use the convention that the larger projectile moves with negative rapidity, which 
implies a positive value for the parameter $R_{ol}^2$). In Fig. \ref{fig:zt} is shown the density of the emission points 
for  charged pions in d-Au collisions of centrality 0-5\%. 
For this centrality  the number of participant nucleons from the Au nucleus is much larger than the two
participants from the deuteron \cite{Bialas:2004su}. The pion rapidity distribution is far from boost invariance and is 
strongly asymmetric in the 
forward-backward direction (Fig. \ref{fig:dndy}). On the large nucleus going side (negative rapidity),
 the fireball is larger and lives longer. The emission time for pions with coordinates in the direction of the larger nucleus 
($z<0$) is larger. As a consequence the moment $G_0=\langle z t\rangle -  \langle z\rangle \langle t \rangle$
is negative, which leads to a nonzero contribution to the $R^2_{ol}$ HBT radius. 

The asymmetric radius $R_{ol}^2$ is larger for  
collisions at lower energies. It is due to the fact that  for central
 d-Au collisions at $200$GeV the slope of the pion distribution
 in rapidity $dN/dy$ at $y=0$ is larger than for p-Pb collisions
 at the LHC (Fig. \ref{fig:dndy}) The time delay  between pions emitted at negative and positive rapidities 
is larger for d-Au than for p-Pb collisions (Fig. \ref {fig:zt}). The distribution of emission points becomes more symmetric for 
pions with larger transverse momentum  and the value of the forward-backward asymmetry in the emission time  decreases.

\section{Azimuthally sensitive HBT analysis}

\label{sec:ahbt}

In the azimuthally sensitive HBT analysis with respect to the second order event plane, the correlation function 
is calculated for pion pairs restricted to bins 
in the azimuthal angle.  The azimuthal angle $\Phi$ 
in the range $[0,\pi]$ is subdivided into 6 bins.  For each freeze-out hypersurface, 
the second order event plane angle $\Psi_2$ is  extracted from
 $N_e=5000$ combined events (each with multiplicity $M_j$)
\begin{equation}
v_2 e^{i2\Psi_2}= \frac{\sum_{j=1}^{N_e}\sum_{l=1}^{M_j} e^{i2\phi_l}}{\sum_{j=1}^{N_e} M_j} \ ,
\end{equation}
 $\phi_l$ are the azimuthal angles of charged particles with $|\eta|<2$.
There is no correction for the  event-plane resolution. On the other hand, the event-plane resolution may constitute
a severe issue for the actual experimental analysis of the azimuthally sensitive HBT  for d-Au collisions, where the multiplicity 
is relatively small.

The Gaussian fit functions for the each azimuthal angle and pair transverse momentum bin involves 
$5$ radii parameters
\begin{equation}
C(q,k_\perp,\Phi)=1+ \lambda  e^{-R_o^2 q_{o}^2-R_s^2q_{s}^2-R_l^2q_{l}^2-2R_{ol}^2 q_{o}q_{l}-2R_{os}^2 q_{o}q_{s}} 
\ . \label{eq:gauss3} 
\end{equation}
The parameter $R^2_{sl}$ is consistent with zero within the accuracy of the fit and is not considered in the analysis.
The azimuthal dependence of the radii is described up to the second
harmonic
\begin{eqnarray}
 R_{o}^2(\Phi)&=&R_{o,0}^2+2R_{o,2}^2\cos(2\Phi) \nonumber \\
 R_{s}^2(\Phi)&=&R_{s,0}^2+2R_{s,2}^2\cos(2\Phi) \nonumber \\
 R_{l}^2(\Phi)&=&R_{l,0}^2+2R_{l,2}^2\cos(2\Phi) \nonumber \\
 R_{os}^2(\Phi)&=&2R_{os,2}^2\sin(2\Phi) \ . \label{eq:rphi}
\end{eqnarray}
For the out-long radius
\begin{equation}
 R_{ol}^2(\Phi)=R_{ol,0}^2+2R_{ol,2}^2\cos(2\Phi)
\end{equation}
the second harmonic term $R^2_{ol,2}$ could not be identified from the fit and in the following the azimuthal dependence
of $R^2_{ol}$ is not considered. The angle independent out-long parameter $R^2_{ol,0}$,
 due to forward-backward asymmetry, is discussed in Sect. \ref{sec:rol}.
 
\begin{figure}
\includegraphics[width=.38\textwidth]{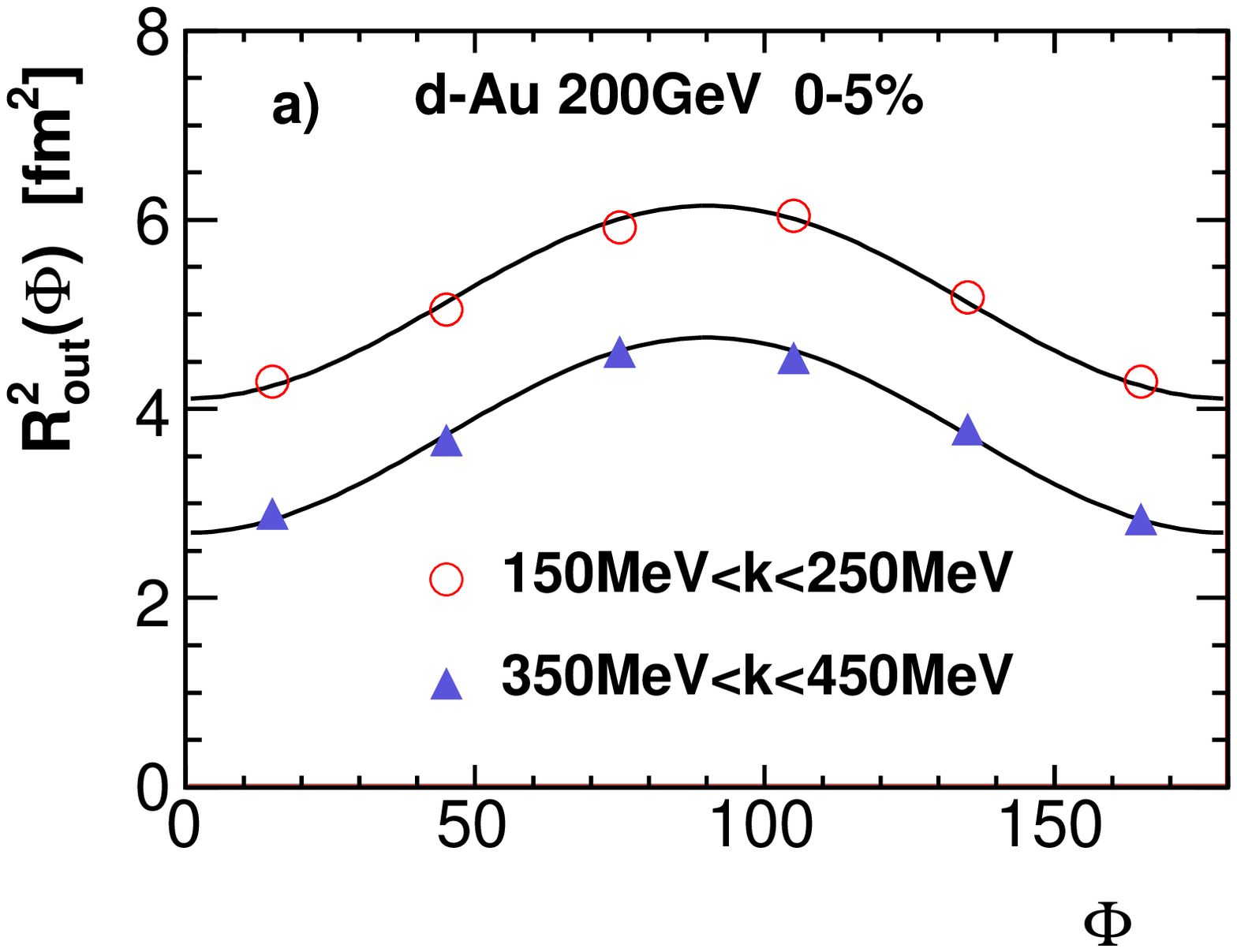}\\
\includegraphics[width=.38\textwidth]{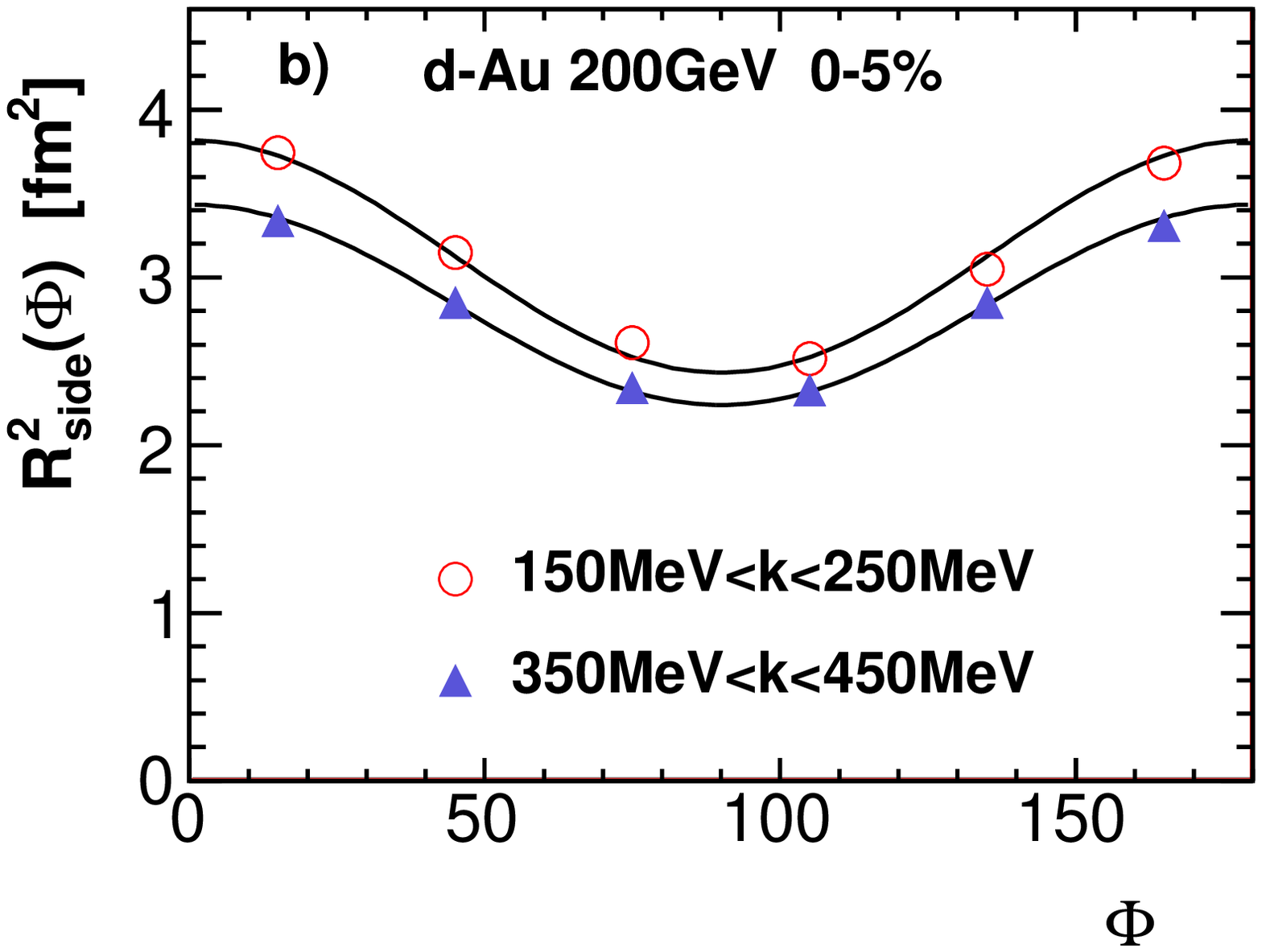}\\
\includegraphics[width=.38\textwidth]{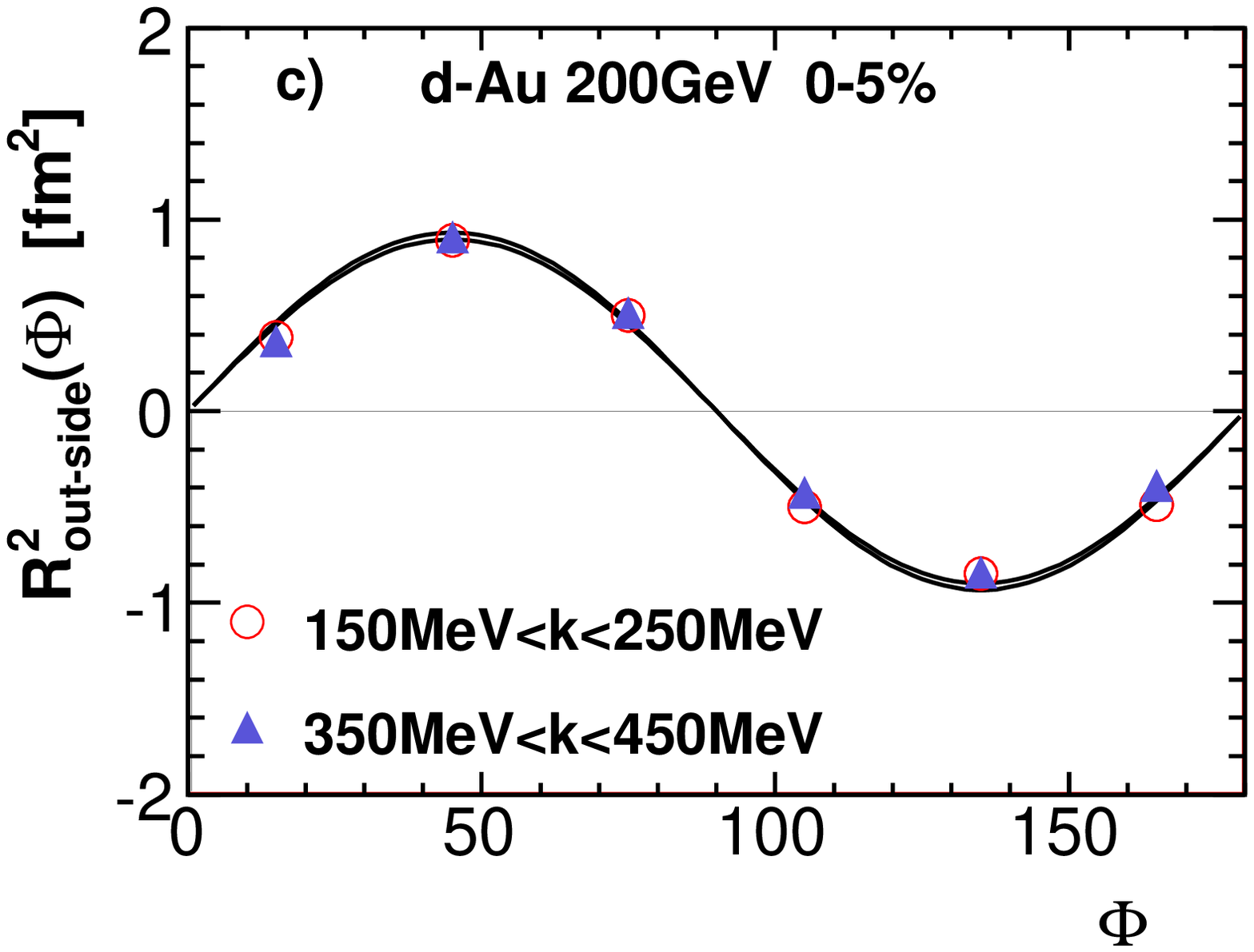}\\
\includegraphics[width=.38\textwidth]{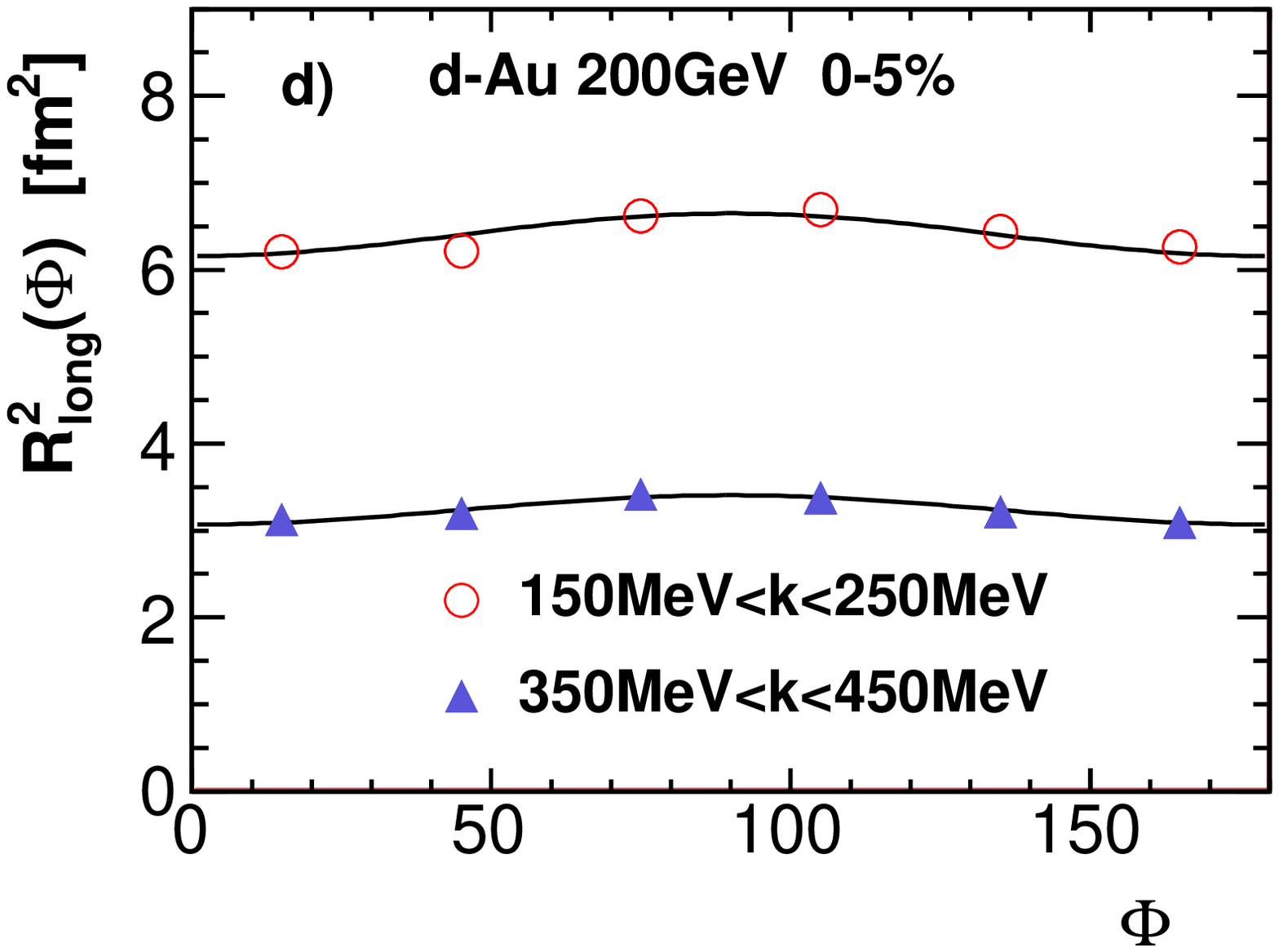}\\
\caption{ (color online)  The azimuthal angle dependence of  the HBT radii $R_{o}$ (panel a),  $R_{s}$ (panel b),   $R_{os}$ (panel c) and $R_{l}$ (panel d) for $150$MeV$<k_\perp<250$MeV (circles) and $350$MeV$<k_\perp<450$MeV (triangles), the solid lines
 represent the second harmonic fit of the angular dependence of the radii (Eq. \ref{eq:rphi}).}
\label{fig:rphi}
\end{figure}

\begin{figure}
\includegraphics[width=.38\textwidth]{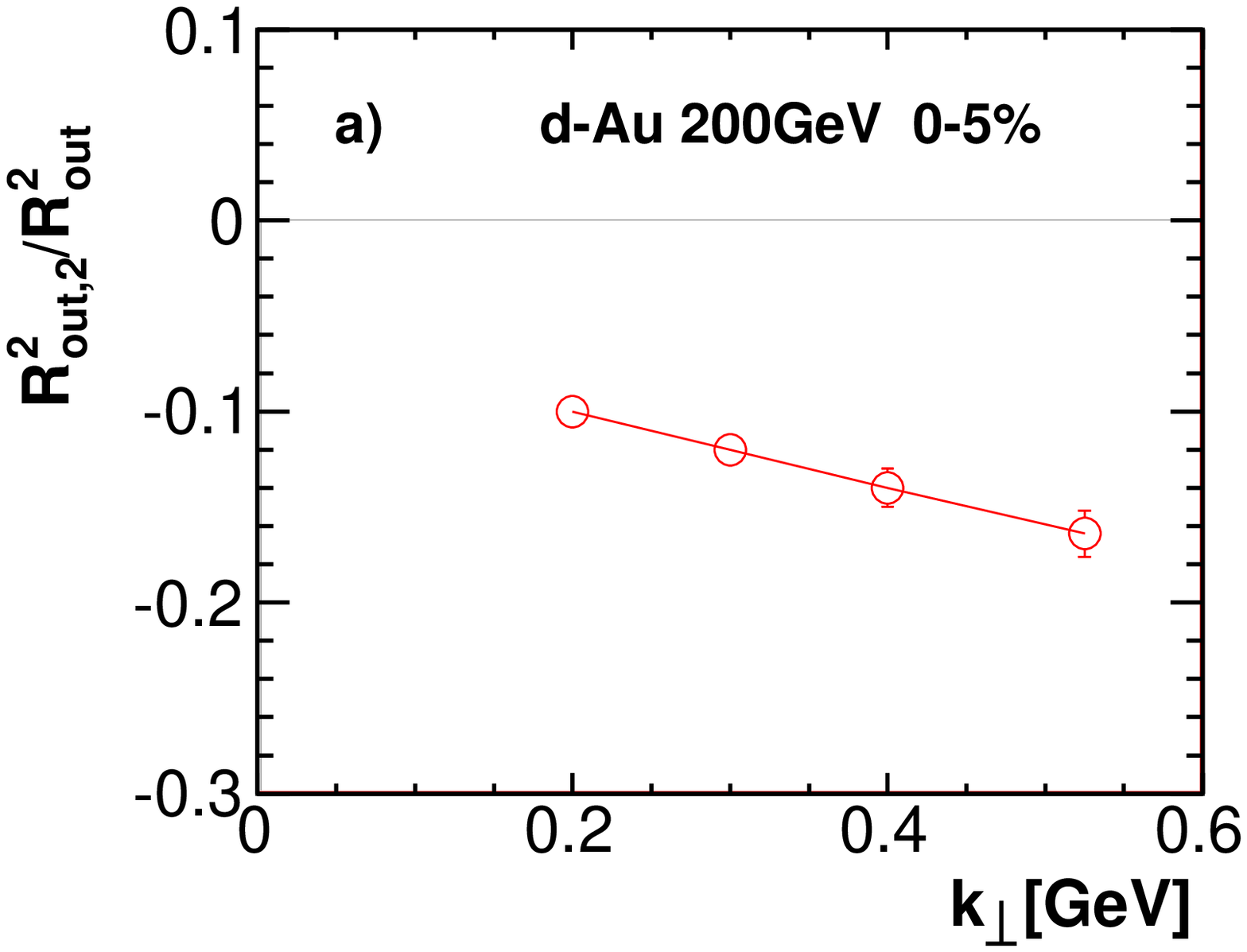}\\
\includegraphics[width=.38\textwidth]{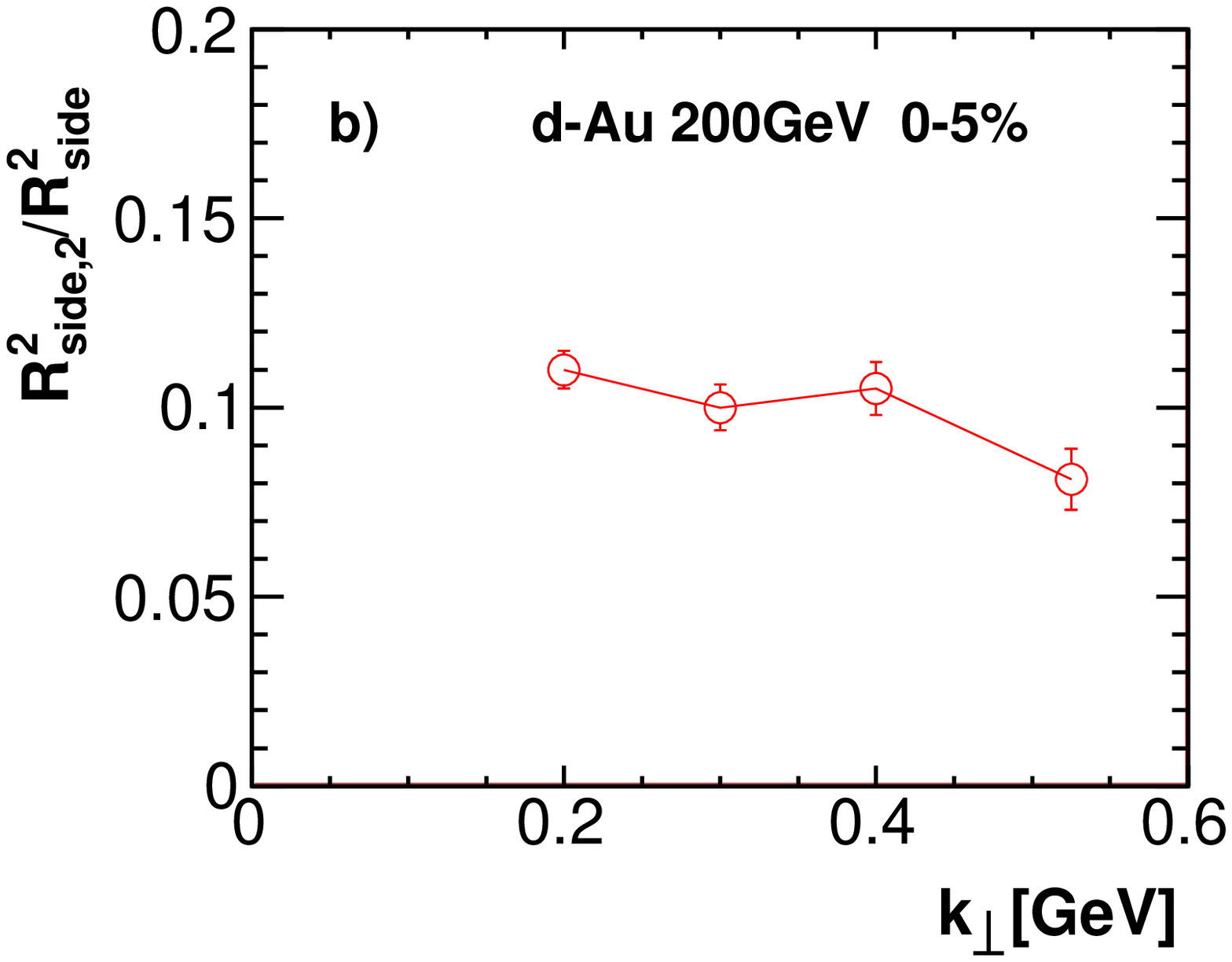}\\
\includegraphics[width=.38\textwidth]{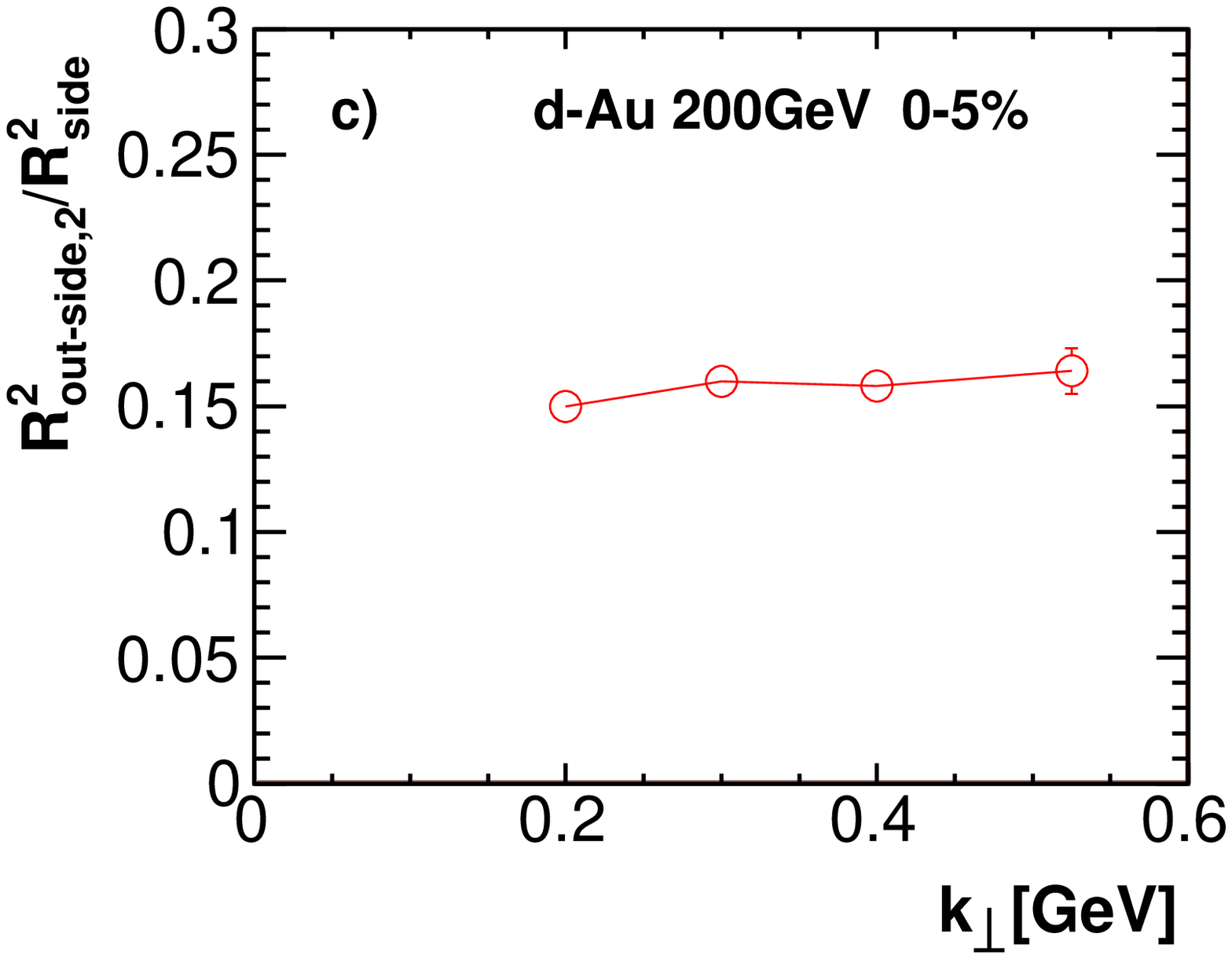}\\
\includegraphics[width=.38\textwidth]{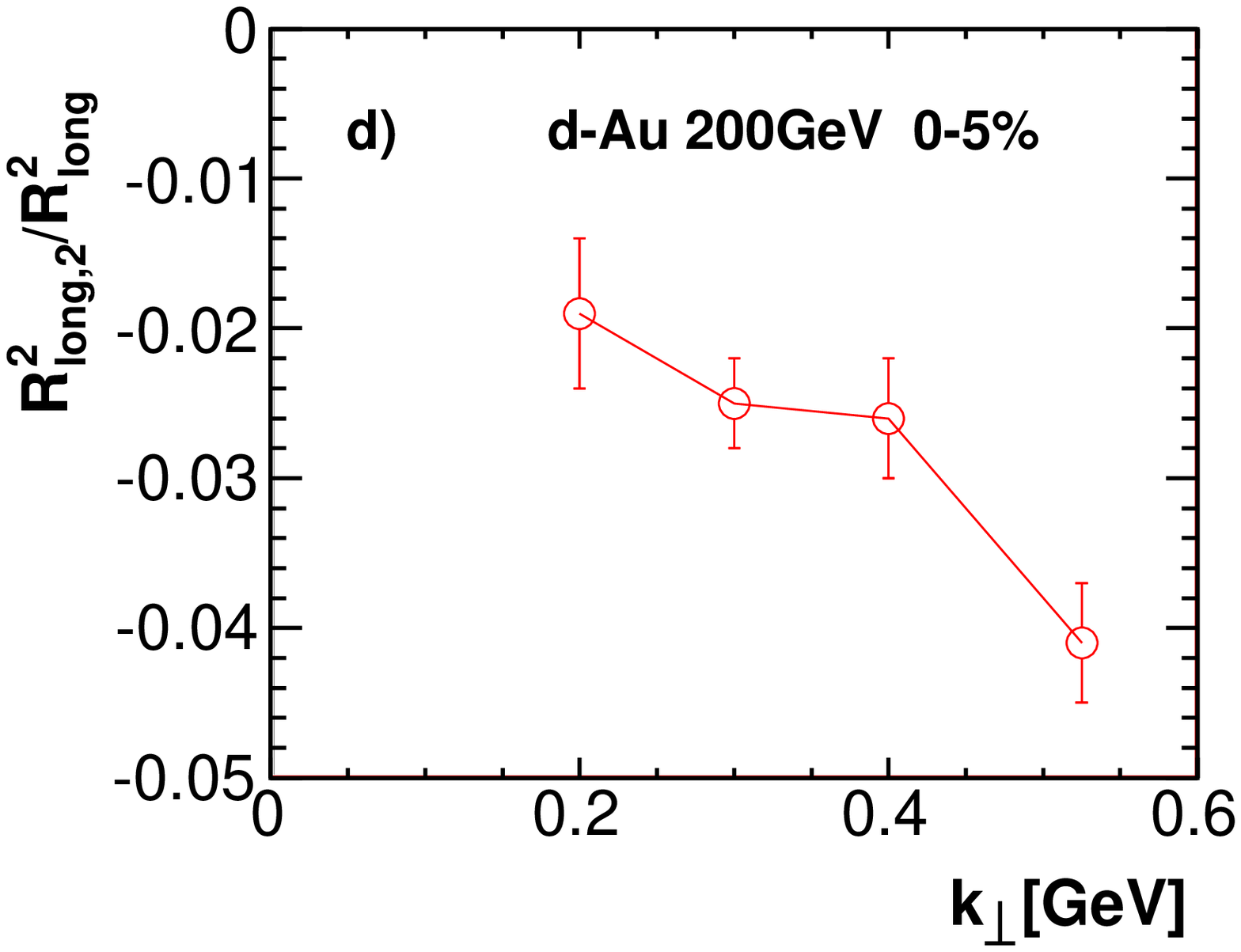}\\
\caption{ Second order Fourier coefficients of the oscillations of the  HBT radii 
 with respect to the second order event plane for d-Au collisions at $200$~GeV,  
$R_{o,2}^2/R_{o,0}^2$ (panel a),  $R_{s,2}^2/R_{s,0}^2$ (panel b),   $R_{os,2}^2/R_{s,0}^2$ (panel c), $R_{l,2}^2/R_{l,0}^2$ (panel d). }
\label{fig:r2}
\end{figure}

The azimuthal angle dependence of the HBT radii for two transverse momentum bins is shown in Fig. \ref{fig:rphi}. The dependence
 can be described using the zeroth and second harmonic  as given by Eq. (\ref{eq:rphi}). The side radius is larger for the  
in plane  direction, for pion pairs emitted in that direction the source size is dominated by the separation in the transverse plane 
 between the proton and the neutron of the incoming deuteron. The fireball is wider when observed in plane, as the deuteron is 
preferably oriented out of plane.
The out radius show the reverse behavior, it is smaller when viewed from the direction with the larger transverse flow.

The second order Fourier coefficients of the HBT radii with respect to the second order event plane are shown in Fig. \ref{fig:r2}.
The second harmonic coefficients are fitted from the relation  (\ref{eq:rphi}), correcting for finite bin width. 
The coefficient $R_{o,2}^2$ is negative, while $R^2_{s,2}$ is positive. 
The dependence on the pair momentum is weak. The large value of $R^2_{s,2}$
reflects the dependence of the geometrical size of the emission source on the angle, in a very similar way as for
 nucleus-nucleus collisions at finite impact parameter \cite{Adams:2003ra,*Adare:2014vax,*Loggins:2014uaa}.

 For d-Au collisions at 0-5\% centrality  the second order harmonic in the azimuthal dependence of the HBT radii is very pronounced. For the strongly
deformed fireball in d-Au collisions, both the shape and the flow at freeze-out have a large quadrupole anisotropy.
The second order harmonic in the $R_{l}$ radius, while it is smaller than for $R_{o}$ and $R_s$,   can be extracted from the
 identical pion correlation function. The asymmetry of the flow generates a contribution proportional to 
 $J_2$ in Eq. (\ref{eq:aradii}), that can be measured. If a similar sign and magnitude of $R^2_{l,2}$ is observed in experiment, it
would indicate that the viscous hydrodynamic model describes well the local flow and the pressure anisotropy at freeze-out.

\section{Summary}

We present a study of  HBT correlations for d-Au collisions at $200$GeV. The experimental analysis of the azimuthal 
correlations  indicates that collective flow appears in such collisions. The interferometry analysis is performed on 
events generated in the viscous hydrodynamic model that describes 
 the observed flow asymmetry \cite{Adare:2013piz}.
The identical pion correlation function is constructed in bins of average transverse momentum of the pion pair. From a Gaussian fit to the correlation function, the  HBT radii  $R_{s}$,
$R_{o}$, and $R_{l}$ are extracted. The  values of the HBT radii are consistent with the 
experimental measurements \cite{Adare:2014vri}.  
The model underestimates the HBT radii for peripheral Au-Au collisions at the same energy. It may indicate that some details
 of the energy deposition in the fireball and of the flow profile are  incorrect for the chosen Glauber model initial conditions, 
further effects may be due to pre-equilibrium flow or non-femtoscopic correlations that are unaccounted for in the model.
For d-Au collisions the most important geometrical scale comes from the r.m.s. radius of the deuteron  and is correctly taken into account in the generation of the initial fireball.

In asymmetric systems, such as d-Au or p-Pb collisions, the fireball  has no forward-backward reflection symmetry 
in rapidity around midrapidity.
The multiplicity as function of pseudorapidity is higher on the  side of the large nucleus. 
In the hydrodynamic model, the fireball lives longer and undergoes a  stronger transverse  
expansion on that side \cite{Bozek:2013sda}. The delay in the emission time 
of pions on the side of the larger nucleus gives observable effects for the interferometry correlations, 
a term combining the out and long directions appears in the Gaussian formula for the correlation function. 
The calculation predicts a nonzero value for the corresponding parameter $R^2_{ol}$ in asymmetric d-Au or p-Pb collisions. 

For central d-Au collisions the system is  strongly asymmetric in azimuthal angle. 
The geometrical size is larger out of plane and the 
transverse flow is stronger in plane. The effects of this asymmetry are clearly visible in azimuthally sensitive HBT analysis with respect to the second order event plane. The three HBT radii show a significant  $\cos(2\Phi)$   angular dependence.
The second harmonic coefficient for the side radius is positive, as expected from the geometry of the emission source.
The reverse is observed for the out direction, $R^2_{o,2}$ is negative.
The strong elliptic flow generates a small angular dependence in the long direction $R_l$.
The  out-side radius $R^2_{os}$ is non-zero and proportional to $\sin(2\Phi)$.

The observed angular correlations in d-Au collisions \cite{Adare:2013piz} can originate from   color glass condensate
effects \cite{Dusling:2013oia} or from the collective flow \cite{Bozek:2011if,Bzdak:2013zma,Qin:2013bha,*Nagle:2013lja}.
The experimental 
observation of the azimuthal dependence of the HBT radii with respect to the second order event plane would mean that  
there is a correlation between the geometrical orientation of the fireball in d-Au collisions and 
the preferred flow direction. 
This would indicate that 
 collective flow is the dominant source of the observed correlations, as the  collective
 expansion translates the initial geometrical asymmetry into the azimuthal momentum asymmetry of the observed particles.

\section*{Acknowledgments}
Supported  by National Science Centre, Grant No.
DEC-2012/05/B/ST2/02528 and by PL-Grid Infrastructure.

\bibliography{../hydr}

%\appendix*

\end{document}